# *Optical Tractor Beam with Chiral Light*

David E. Fernandes, Mário G. Silveirinha[1]

*University of Coimbra, Department of Electrical Engineering – Instituto de Telecomunicações, 3030-290, Coimbra, Portugal,*

*dfernandes@co.it.pt, mario.silveirinha@co.it.pt*

**Abstract**

We suggest a novel mechanism to induce the motion of a chiral material body towards an optical source with no optical traps. Our solution is based on the interference between a chiral light beam and its reflection on an opaque mirror. Surprisingly, it is theoretically shown that the electromagnetic response of the material may be tailored in such a way that independent of the specific body location with the respect to the mirror, it is always pulled upstream against the photon flow associated with the incoming wave. Moreover, it is proven that by controlling the handedness of the incoming light it may be possible to harness the sign of the optical force, switching from a pulling force to a pushing force.

[1] To whom correspondence should be addressed: E-mail: mario.silveirinha@co.it.pt

The interaction between light and matter is manifested macroscopically in terms of a radiation pressure that may be used to trap micrometer-sized particles and neutral atoms, and for atom cooling [1-5]. Optical tweezers [6] are commonly used in molecule manipulation for biological and physical chemistry applications [7-8]. These advances were made possible by the advent of lasers, which provide the means to effectively boost the usually weak optomechanical interactions.

Recently, there has been a great interest in new techniques for optical manipulation of nanoparticles using unstructured (gradientless) light. Intuitively, one might expect that when a uniform light beam impinges on a polarizable body, the translational force acting on the body should always push it towards the direction of propagation. Notably, several groups demonstrated theoretically, and in some cases also experimentally, that travelling light beams may allow pulling micron-sized particles towards the light source [9-25]. Such solutions are known as optical "tractor beams", which may be defined as engineered light beams that exert a negative scattering force on a polarizable body, forcing it to move along a direction opposite to the photon flow. A pulling force can be obtained with interfering non-diffractive beams [10], based on optical-conveyor belts [9], using structured media supporting backward waves [12], with the help non-paraxial Bessel beams [13, 14, 16, 24], and using gain media [22, 23].

In this work, we put forward a different paradigm for a tractor beam based on unstructured light and suitably designed chiral metamaterials. It is shown that by controlling the polarization state helicity it is possible to switch from a pushing force (downstream motion) to a pulling force (upstream motion). The emergence of pulling forces resulting from the interaction of chiral particles with light Bessel beams has been recently discussed in [24, 26]. Our proposal to harness the sign of the optical force is based on entirely different physical mechanisms, and does not require complex Bessel beams.

For the sake of clarity, we consider a simple canonical geometry wherein the body of interest is a planar slab with thickness $d$ (Fig. 1). Nevertheless, the ideas that follow can be readily generalized to other geometries, e.g. to small nanoparticles. Our solution relies on the interference of the downstream wave emitted by a light source with an upstream wave created by reflection on an opaque mirror (Fig. 1). It is well known that the interference of the two waves creates a standing wave, and that usually a material body is pushed to a spot wherein the electric field intensity reaches a local maximum. Surprisingly, we show that the material response may engineered such that *independent* of the location of the slab with respect to the mirror, the body is always pulled upstream towards the light source. Our solution uses unstructured light (consistent with the usual definition of a tractor beam) and explores the fact that in most scenarios of interest the relevant objects stand above some surface (the mirror). We fully take into account and explore to our advantage the presence of such a mirror.

It is a simple exercise to show that when a slab stands alone in free-space and is illuminated by a plane wave propagating along the $z$-direction the time-averaged translational optical force is $F_{z,\text{av}} / A = W_{\text{av}}^{inc}\left(1+|R|^2-|T|^2\right)$, where $A$ is the area of the slab cross-section, $W_{\text{av}}^{inc}=\frac{1}{4}\left(\varepsilon_0\left|\mathbf{E}^{inc}\right|^2+\mu_0\left|\mathbf{H}^{inc}\right|^2\right)$ is the time-averaged energy density of the

incident plane wave, and $R$ and $T$ are the reflection and transmission coefficients. This formula is obtained by noting that in a stationary time-harmonic regime ($\mathbf{E}(\mathbf{r},t)=\mathrm{Re}\{\mathbf{E}(\mathbf{r})e^{-i\omega t}\}$) the time-averaged force $\mathbf{F}_V$ is given by $\mathbf{F}_V = \int_{\partial V} dV\, \hat{\mathbf{n}}\cdot \mathrm{Re}\{\underline{\mathbf{T}_c}\}$ where $\underline{\mathbf{T}_c}$ is the (complex-valued) Maxwell stress tensor $\underline{\mathbf{T}_c}=\frac{1}{2}\left[\varepsilon_0 \mathbf{E}\otimes\mathbf{E}^* - \frac{1}{2}\varepsilon_0\mathbf{E}\cdot\mathbf{E}^*\,\underline{\mathbf{1}}+\frac{1}{\mu_0}\mathbf{B}\otimes\mathbf{B}^* - \frac{1}{2\mu_0}\mathbf{B}\cdot\mathbf{B}^*\,\underline{\mathbf{1}}\right]$, $\underline{\mathbf{1}}$ is the identity tensor, $\partial V$ is the boundary surface of the body with volume $V$, and $\hat{\mathbf{n}}$ is the outward unit normal vector [27, 28]. For passive materials the conservation of energy implies that the optical force is non-negative, $F_{z,\mathrm{av}} \geq 0$, and that the body is dragged downstream. In particular, the optical force is minimal ($F_{z,\mathrm{av}} = 0$) when the material body is transparent to the incoming wave ($|R|=0$ and $|T|=1$). Interestingly, for active materials the transmission coefficient can be greater than unity and hence it is possible to have $1+|R|^2-|T|^2<0$ and a pulling (negative) force [22, 23].

In presence of the mirror wall, the material body interacts with both the downstream and upstream waves. The above discussion suggests that to have a negative optical force the downstream wave should desirably be perfectly transmitted through the material, so that no momentum is imparted on the body by this wave. On the other hand, the upstream wave should be either absorbed or reflected by the material body so that it can transfer a negative momentum to the body, and induce a negative optical force. Thus, the material body must scatter the waves propagating upstream and downstream in an *asymmetric* manner. Clearly, this cannot be achieved with a standard dielectric slab because by symmetry, the scattering response is the same for the two waves. Moreover, a stack of conventional dielectrics is also not a valid option because if $|T|=1$ for the downstream wave, then, because of the Lorentz reciprocity theorem [27], $|T|$ is also required to be identically one for the upstream wave.

Notably, chiral metamaterials provide a way out of this bottleneck, and enable one to engineer the scattering properties such that the transmission of the upstream and downstream waves may be strongly asymmetric. Chiral media are a special class of materials with magneto-electric coupling determined by the handedness of the structural unities [29]. Chiral materials endow us with the means to achieve polarization rotation (optical activity) [30-32], polarization conversion [33], and negative refraction [34-37]. The constitutive relations of a chiral material are $\mathbf{D}=\varepsilon_0\varepsilon\mathbf{E}+i\chi\sqrt{\mu_0\varepsilon_0}\mathbf{H}$ and $\mathbf{B}=-i\chi\sqrt{\mu_0\varepsilon_0}\mathbf{E}+\mu_0\mu\mathbf{H}$, where $\chi$ is the dimensionless chirality parameter, $\varepsilon$ is the relative permittivity, and $\mu$ is the relative permeability. The photonic modes in an unbounded chiral medium are right circular polarized (RCP) waves or left circular polarized (LCP) waves that are characterized by distinct refractive indices $n_\pm=\sqrt{\mu\varepsilon}\pm\chi$, where $n_+$ ($n_-$) stands for the refractive index of RCP (LCP) waves propagating along the +$z$-direction [29]. It is well known that a chiral slab scatters RCP and LCP waves differently, such that the transmission and reflection coefficients for circularly polarized waves are [29, 38]:

$$T_\pm=\frac{4ze^{ik_\pm d}}{(1+z)^2-(1-z)^2e^{2ink_0d}}, \qquad R\equiv R_\pm=\frac{(1-z^2)\left(e^{2ink_0d}-1\right)}{(1+z)^2-(1-z)^2e^{2ink_0d}}. \qquad (1)$$

In the above, $d$ is the thickness of the chiral slab, $k_{\pm} = k_0 \left( n \pm \chi \right)$, $k_0 = \omega / c$, $n = \sqrt{\mu \varepsilon}$, and $z = \mu / n = \sqrt{\mu / \varepsilon}$ is the normalized impedance. Hence, the reflection coefficient is identical for the two polarization states but, crucially, the transmission coefficient is polarization dependent. Now, the key observation is that in presence of an opaque mirror, e.g. for a metallic mirror, the wave polarization state is switched upon reflection in the mirror, such that if the downstream wave has RCP (LCP) polarization then the upstream wave has LCP (RCP) polarization. This implies that the scattering of the upstream and downstream waves can be strongly asymmetric because the transmission coefficient for RCP waves ($T_+$) differs from the transmission coefficient for LCP waves ($T_-$). The asymmetric transmission of electromagnetic waves through chiral metamaterials was discussed by different groups in other contexts [39-41]. For future reference, let us denote $T_{21}$ as the transmission coefficient for a circularly polarized wave propagating downstream that impinges on a chiral slab standing alone in free-space ($T_{21} = T_{\pm}$ for an incident RCP/LCP wave; the subscript "21" indicates that the incident wave propagates from the region 1 to the region 2), and $T_{12}$ as the transmission coefficient for the corresponding upstream wave ($T_{12} = T_{\mp}$). One important remark is that the conservation of energy and the fact that $R$ is polarization independent, imply that $\left| T_{12} \right| = \left| T_{21} \right|$ in the absence of material absorption. Thus, a strong asymmetric transmission *requires* some material loss.

To investigate the opportunities created by a chiral response in the context of optical manipulation, next we derive an explicit formula for the optical force. The electromagnetic fields in the scenario of Fig. 1 are of the form:

$$\mathbf{E}(z)=\left[C^{+}(z)+C^{-}(z)\right]\mathbf{e}_{\pm}, \qquad \eta_0\mathbf{H}(z)=-\left[C^{+}(z)-C^{-}(z)\right](\pm i)\mathbf{e}_{\pm}, \tag{2}$$

with $\mathbf{e}_{\pm}=(\hat{\mathbf{x}}\pm i\hat{\mathbf{y}})/\sqrt{2}$, where the subscript $\pm$ determines if the incoming (downstream) plane wave has RCP (+) or LCP (–) polarization. The function $C^{+}(z)$ ($C^{-}(z)$) represents the complex amplitude of the downstream (upstream) wave at each point of space. Using the Maxwell-stress tensor, it can be readily found that the translational force acting on the chiral slab is

$$\frac{F_{z.\mathrm{av}}}{A}=-\frac{\varepsilon_0}{4}\left(\left|C^{+}(z)+C^{-}(z)\right|^2+\left|C^{+}(z)-C^{-}(z)\right|^2\right)\Bigg|_{z=z_{in}}^{z=z_{out}}, \tag{3}$$

where $z=z_{in}$ ($z=z_{out}$) are the coordinates of the input (output) interfaces of the slab (Fig. 1). After some simplifications, the force can be written as:

$$\frac{F_{z.\mathrm{av}}}{A}=-\frac{\varepsilon_0}{2}\left(\left|C_{out}^{+}\right|^2+\left|C_{out}^{-}\right|^2-\left|C_{in}^{+}\right|^2-\left|C_{in}^{-}\right|^2\right), \tag{4}$$

where $C_{in}^{\pm}=C^{\pm}(z_{in})$ and $C_{out}^{\pm}=C^{\pm}(z_{out})$. Note that $C_{in}^{+}$ is the complex amplitude of the incident electric field at the chiral slab input interface. The coefficients $C_{in}^{\pm}$ and $C_{out}^{\pm}$ are linked by $C_{in}^{-}=R\,C_{in}^{+}+T_{12}\,C_{out}^{-}$ and $C_{out}^{+}=T_{21}\,C_{in}^{+}+R\,C_{out}^{-}$, where $T_{12}$, $T_{21}$, and $R$ are the transmission/reflection coefficients introduced previously for a slab standing alone in free-space. Moreover, when a metallic mirror is placed at a distance $l$ from the slab the coefficients $C_{out}^{\pm}$ are related by $C_{out}^{-}=-e^{i2k_0 l}C_{out}^{+}$. A similar expression is obtained for other mirror types (e.g. with a photonic crystal wall). Putting these results together, it is possible to express $C_{in}^{-}$ and $C_{out}^{\pm}$ as a function of the incident wave complex amplitude $C_{in}^{+}$ as follows:

$$C_{in}^{-}=\left(R-\frac{e^{i2k_0 l}T_{21}T_{12}}{1+Re^{i2k_0 l}}\right)C_{in}^{+},\qquad C_{out}^{+}=\frac{T_{21}}{1+Re^{i2k_0 l}}C_{in}^{+},\qquad C_{out}^{-}=\frac{-e^{i2k_0 l}T_{21}}{1+Re^{i2k_0 l}}C_{in}^{+}. \quad (5)$$

Substituting the above formulas into Eq. (4), we finally find that the optical force satisfies:

$$\frac{F_{z.\mathrm{av}}}{A}=W_{\mathrm{av}}^{inc}\left(1+\left|R-\frac{e^{i2k_0 l}T_{21}T_{12}}{1+Re^{i2k_0 l}}\right|^2-2\left|\frac{T_{21}}{1+Re^{i2k_0 l}}\right|^2\right), \quad (6)$$

where $W_{\mathrm{av}}^{inc}=\frac{\varepsilon_0}{2}\left|C_{in}^{+}\right|^2$ is the time-averaged energy density of the incident wave.

Let us now consider the scenario wherein $|R|=0$ and $|T_{21}|=1$ so that the slab is transparent to the downstream wave and the downstream light-matter interactions are minimized. In this ideal case, the optical force (6) becomes $F_{z.\mathrm{av}}/A=W_{\mathrm{av}}^{inc}\left(|T_{12}|^2-1\right)<0$, i.e. consistent with our intuition the optomechanical interactions cause the material slab to move upstream towards the light source, *independent* of its location with respect to the mirror. Interestingly, the amplitude of the negative force has the maximum value when $|T_{12}|=0$. A simple picture of this scenario [see Fig. 1a] is that initially the downstream beam overtakes the chiral slab without any momentum transfer, and then, upon reflection in the mirror, it is totally absorbed by the chiral slab, transferring in this way its electromagnetic momentum to the slab. Clearly, this regime requires a strongly asymmetric scattering by the chiral slab (ideally, $|R|=0$, $|T_{21}|=1$, $|T_{12}|=0$). It should be noted that the negative momentum is indirectly provided by the mirror wall, which upon reflection flips the sign of the incident radiation momentum. It is important to highlight that the proposed regime is fundamentally different from the conventional light-matter interactions in presence of standing waves, which tend to drag a polarizable object to a field intensity maximum.

A remarkable property of our system is that if the handedness of the downstream light is reversed (e.g. from LCP to RCP) then $T_{21}$ and $T_{12}$ are interchanged. Hence, if for a certain light polarization the downstream light wave originates a negative optical force (e.g. $|R|=0$, $|T_{21}|=1$, $|T_{12}|=0$), then a downstream light wave with opposite handedness originates a positive optical force (e.g. $|R|=0$, $|T_{21}|=0$, $|T_{12}|=1$). Indeed, in the latter situation the downstream light is completely absorbed by the chiral slab never reaching the opaque mirror [see Fig. 1b], and thus it imparts a positive momentum to the material body. Therefore, by controlling the handedness of the downstream light it may be possible to transport a material particle at will, either upstream or downstream with no optical traps, similar to an optical "conveyer belt".

To demonstrate these ideas, we designed a chiral metamaterial such that the meta-atoms are conjugated-gammadions as illustrated in Fig. 2a-b. Previous studies have shown that this metamaterial is characterized by a strong circular dichroism and large optical activity in the microwave and low-THz regime [31, 32]. Evidently the response of this metamaterial is not isotropic, but the key requirement to have a pulling force is the asymmetric light scattering, rather than the isotropy.

To take advantage of high-power laser sources, the meta-atom was engineered to operate around the wavelength $\lambda_0 = 1.55\,\mu\text{m}$. The conjugated-gammadions are made of silver, whose permittivity is described by a Drude dispersion model $\varepsilon_m(\omega) = 1 - \omega_p^2 / \omega(\omega + i\Gamma)$, with parameters $\omega_p / 2\pi = 2175\,[\text{THz}]$ and $\Gamma / 2\pi = 4.35\,[\text{THz}]$, consistent with experimental data reported in the literature [42]. The gammadions are embedded in polyimide with $\varepsilon_p / \varepsilon_0 = 6.25(1 + i0.03)$ [31], which is encapsulated between two layers of silicon ($\varepsilon_s / \varepsilon_0 = 11.9(1 + i0.004)$), as shown in Figure 2b. The meta-atoms are arranged in a periodic square lattice with a deeply subwavelength period $a = 202.4\,\text{nm} \approx 0.131\lambda_0$.

Using the full-wave electromagnetic simulator CST-MWS [43] we numerically obtained the amplitude and phase of the transmission and reflection coefficients for incident circularly polarized light (see Fig. 2c-d). The details of the calculation are given in the supplementary materials [44]. As expected, the reflection coefficients for RCP and LCP waves are equal. Most importantly, the simulations show that the chiral metamaterial has a giant optical activity near $160\,\text{THz}$, $196\,\text{THz}$ and $245\,\text{THz}$, where the scattering of the RCP and LCP waves is strongly asymmetric. This is highlighted in Fig. 2e which depicts the circular dichroism of the structure. Notably, near $196\,\text{THz}$, there is a spectral region wherein $|T_-|^2 - |T_+|^2$ is near to one while the reflection coefficient is near zero, which are exactly the conditions required to have a pulling force with an LCP excitation. In the other spectral regions the asymmetric response is not strong enough to have $\left|T_{-}\right|^2 - \left|T_{+}\right|^2 \approx \pm 1$, as required by a pulling force. Figure 3 depicts the normalized optical force $F_{z.\text{av}} / A W_{\text{av}}^{inc}$ calculated for two particular positions of the slab, $l = 200\text{nm}$ (dashed curves) and $l = 350\text{nm}$ (solid curves) for a downstream wave with either RCP (blue curves) or LCP (green curves) polarization. The results confirm that only in a frequency range around $196\,\text{THz}$, the force exerted by an LCP excitation is negative for the two positions of the PEC screen. Moreover we see that in the same frequency range a downstream RCP wave exerts a positive force on the chiral metamaterial, in agreement with our theory.

Figure 4a and 4c show the optical forces created by RCP and LCP downstream excitations at two fixed frequencies $190\,\text{THz}$ and $196\,\text{THz}$, respectively. As seen, while at $190\,\text{THz}$ –due to the interference of the upstream and downstream waves– the optical force sign depends on the specific position of the slab relative to the mirror, at $196\,\text{THz}$ an LCP excitation always exerts a negative force on the slab whereas an RCP excitation always pushes the slab towards the mirror, independent of the distance $l$. Thus, the LCP excitation mimics, indeed, an optical tractor beam. Here, it is interesting to note that the force varies periodically with the distance between the chiral slab and the metallic screen, with a period $l = \lambda_0 / 2$, where $\lambda_0$ is the light wavelength.

Because the optical force only depends on the position of the slab relative to the mirror, we can define a potential energy as $V = -\int F_{z,\text{av}} dz$, which is depicted in Fig. 4b and 4d for $190\,\text{THz}$ and $196\,\text{THz}$, respectively. As seen, in the former case there are potential wells wherein the material body will inevitably become “trapped”. Quite differently, at $196\,\text{THz}$ the potential energy has a monotonic dependence with $l$ and does not have stationary points so that optical traps are not formed. Hence, depending on the light polarization, the material body is steadily pushed either downstream (RCP polarization) or upstream (LCP polarization). Interestingly, this optical conveyer belt operation may occur in a spectral range from $194.42\,\text{THz}$ to $198.57\,\text{THz}$, which corresponds to a bandwidth of $4.15\,\text{THz}$. Finally, we note that high-power lasers operating at $\lambda_0 = 1.55\,\mu\text{m}$ may generate a light intensity (under continuous wave operation) $S^{inc} \approx 20\,\text{GW/m}^2$ [45]. In this case, an optical force with amplitude $\left|F_{z.\text{av}} / A W_{\text{av}}^{inc}\right| = 1.0$ corresponds to an optical pressure of the order $\left|F_{z.\text{av}} / A\right| = S^{inc} / c \approx 70\,\text{Pa}$, which is a large value at the nanoscale [7] and which we estimate to be at least 3500 larger than the gravitational pressure exerted on the chiral metamaterial. In the supplementary materials [44], we contrast the unique optical manipulations made possible by the asymmetric response of the chiral metamaterial with what is obtained using standard dielectrics.

In summary, it was theoretically demonstrated that by using circularly polarized light and a planar chiral metamaterial with a tailored response it may be possible to mimic an optical tractor beam. By controlling the incident light polarization one may switch from a pushing force to a pulling force, independent of the specific position of chiral slab. We envision that our findings may lead into new inroads in the optical manipulation of micro and nanoparticles with no optical traps.

**Acknowledgements**

This work was funded by Fundação para Ciência e a Tecnologia under project PTDC/EEI-TEL/2764/2012. D. E. Fernandes acknowledges support by Fundação para a Ciência e a Tecnologia, Programa Operacional Potencial Humano/POPH, and the cofinancing of Fundo Social Europeu under the fellowship SFRH/BD/70893/2010.

# Figures:

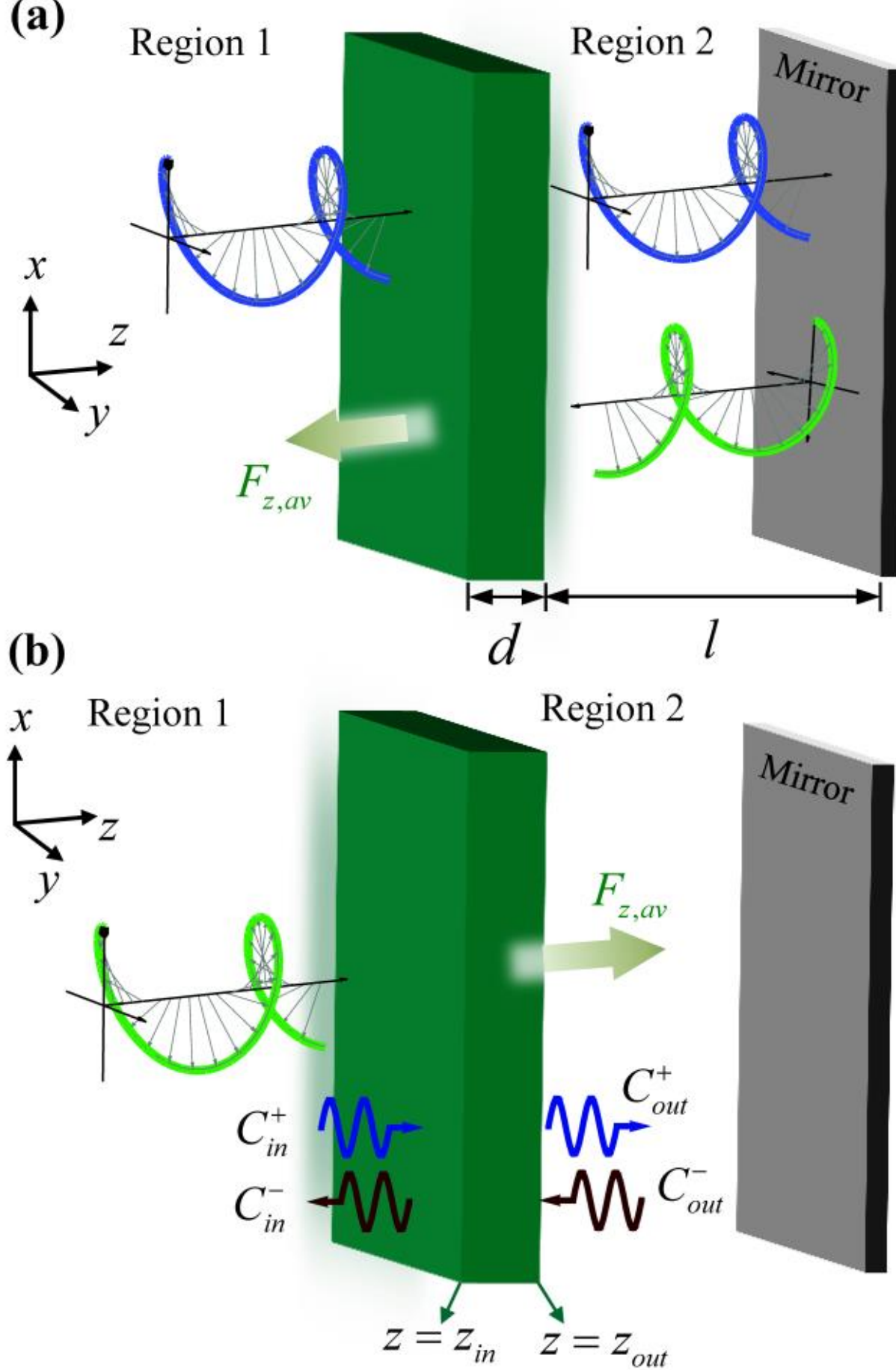

Fig. 1 (color online) A chiral material slab with thickness $d$ is illuminated by a normal incident circularly polarized plane wave. A mirror is placed at a distance $l$ ahead of the chiral slab. **(a)** An incoming beam with a specific handedness overtakes the chiral slab without any momentum transfer and upon reflection in the mirror is totally absorbed by the slab originating a pulling force *independent* of the distance $l$. **(b)** An incoming beam with the opposite handedness is totally absorbed by the chiral slab originating a pushing force.

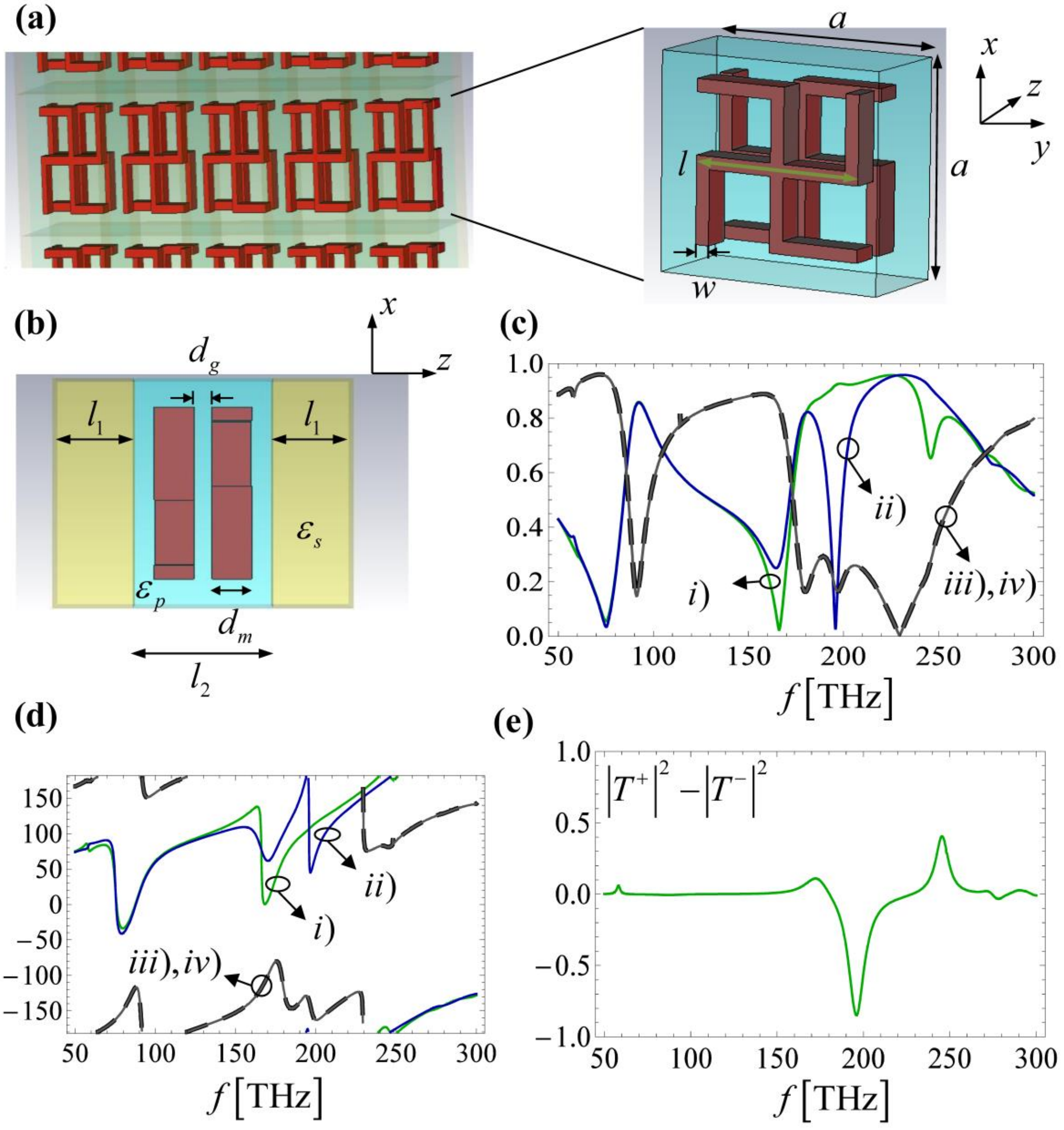

Fig. 2 (color online) **(a)** Geometry of the planar chiral metamaterial. The length of the central arm of the gammadions is $l = 151.8\,\text{nm}$, the width is $w = 12.2\,\text{nm}$. The meta-atom is arranged in a periodic square lattice with period $a = 202.4\,\text{nm}$. **(b)** Side view of the meta-atom: the conjugated-gammadions are made of silver with thickness $d_m = 25\,\text{nm}$ and are separated by a distance $d_g = 11.6\,\text{nm}$. The two gammadions are embedded in a polyimide slab with thickness $l_2 = 88.4\text{nm}$ and permittivity $\varepsilon_p / \varepsilon_0 = 6.25(1 + i0.03)$. Two layers of silicon, with permittivity $\varepsilon_s / \varepsilon_0 = 11.9(1 + i0.004)$ and thickness $l_1 = 50.6\,\text{nm}$, are placed on each side of the polyimide material. The total thickness of the meta-atom is $d = 2l_1 + l_2 = 189.6\,\text{nm} \approx 0.122\lambda_0$. **(c)**

Amplitude of: *i)* transmission coefficient for incident LCP waves, *ii)* transmission coefficient for incident RCP waves, *iii)* reflection coefficient for incident LCP waves, *iv)* reflection coefficient for incident RCP waves. **(d)** Similar to **(c)** but for the phase of the transmission and reflection coefficients. **(e)** Circular dichroism.

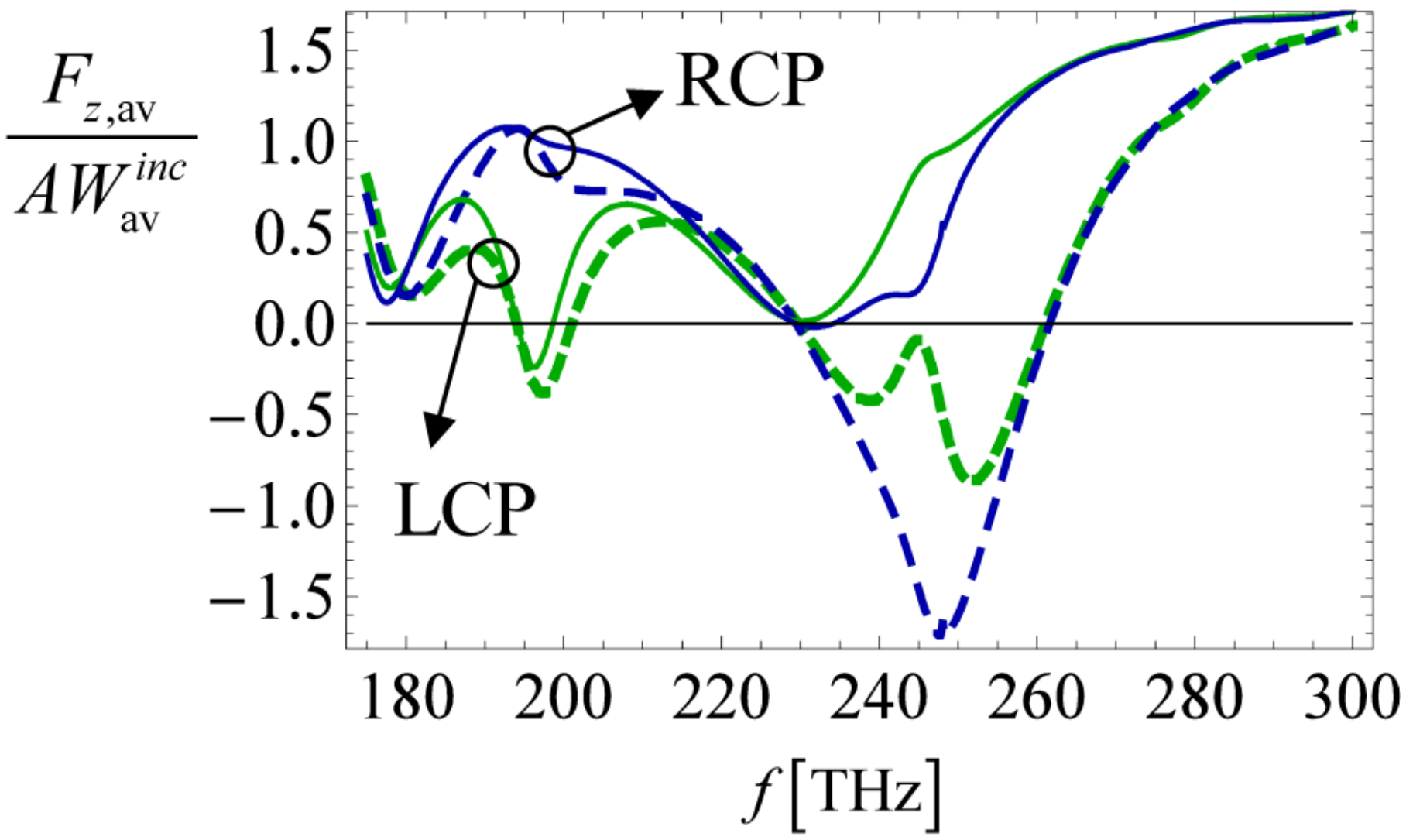

Fig. 3 (color online) Normalized force $F_{z.\mathrm{av}}/AW_{\mathrm{av}}^{inc}$ as a function of the frequency for two fixed positions of the chiral slab: $l = 200\mathrm{nm}$ (dashed curves) and $l = 350\mathrm{nm}$ (solid curves) for LCP downstream light (green curves) and RCP downstream light (blue curves).

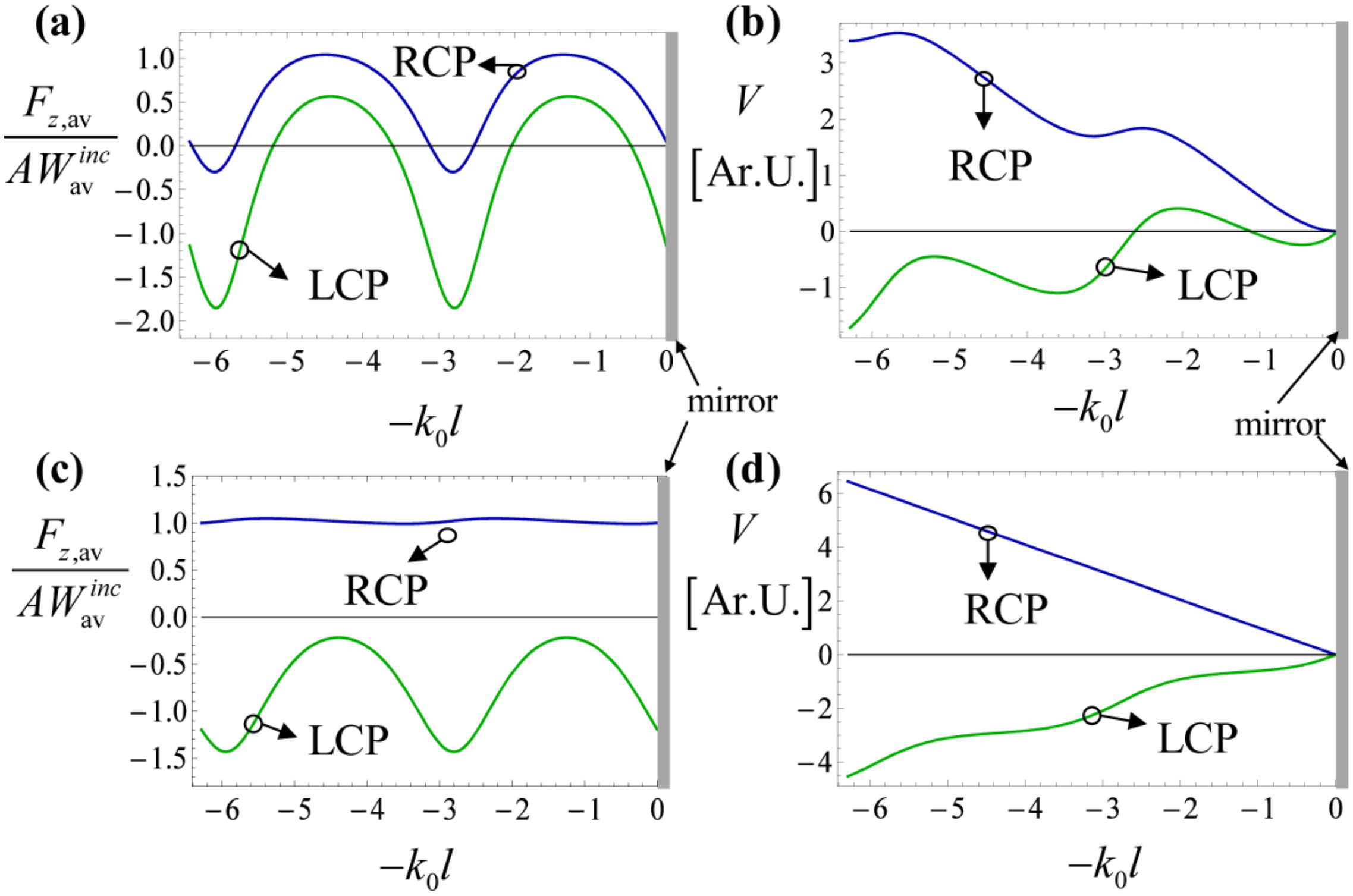

Fig. 4 (color online) **(a)** Normalized force $F_{z.\text{av}}/AW_{\text{av}}^{inc}$ and **(b)** Potential (in arbitrary units) as a function of the normalized distance $k_0 l$ between the chiral slab and the PEC screen at a fixed frequency $f = 190\,\text{THz}$ for LCP downstream light (green curves) and RCP downstream light (blue curves). **(c)** and **(d)** Similar to **(a)** and **(b)**, respectively, but for a fixed frequency $f = 196\,\text{THz}$.